\documentstyle[aps,eqsecnum,preprint,prd,epsf]{revtex}
\begin{document}
\addtolength{\jot}{10pt}
\tighten
\draft
\preprint{\vbox{\hbox{BARI-TH/95-219 \hfill}
                \hbox{UTS-DFT-95-13 \hfill}
                \hbox{December 1995 \hfill}
                 }}
\vskip 2cm
\title{\bf On the $b$ Quark Kinetic Energy in the $\Lambda_b$ \\}
\vskip 1cm
\author{P. Colangelo $^{a}$, C.A. Dominguez $^{b}$, G. Nardulli $^{a,c}$ and N.
Paver $^d$\\ \vskip 1cm}

\address{
$^{a}$ Istituto Nazionale di Fisica Nucleare, Sezione di Bari, Italy\\
  \vspace{2mm}
$^{b}$ Physics Department, University of Capetown, South Africa\\
  \vspace{2mm}
$^{c}$ Dipartimento di Fisica, Universit\'a di Bari, Italy \\
  \vspace{2mm}
$^{d}$ Dipartimento di Fisica Teorica, Universit\'a 
di Trieste, and  \\
  \vspace{2mm}
Istituto Nazionale di Fisica Nucleare, Sezione di Trieste, Italy\\}
\maketitle
\vskip 0.5cm
\begin{abstract}
We consider the problem of determining the matrix element of the  heavy quark
kinetic energy operator $\cal K$
between $\Lambda_b$ baryonic  states
by three point function QCD sum rules in the Heavy Quark Effective Theory. 
We discuss the main features  and the uncertainties associated to such 
an approach.
Numerically, we find the result:
${\langle{\cal K}\rangle}_{\Lambda_b} = - (0.6 \pm 0.1) \; GeV^2$.
Within the uncertainty of the method, such result indicates a 
value of the heavy quark kinetic energy in the baryon 
similar in size to that in the meson.
\end{abstract}
\vskip 1.cm


\clearpage

\section{Introduction}
Important progresses in the theoretical description of hadrons containing 
one heavy quark have been allowed by the development of the Heavy Quark 
Effective Theory (HQET) \cite{hqet}. Based on the spin-flavor symmetry of QCD, 
valid in the limit of infinite heavy quark mass $m_Q$, this 
framework provides a systematic expansion of heavy hadron spectra and 
transition amplitudes in terms of the leading contribution (for 
$m_Q\to\infty$), plus corrections decreasing as powers of $1/m_Q$. Being 
derived from `first principles', i.e. from general properties of QCD, 
such an approach has the advantage of being model-independent to a large 
extent, and predictive at the same time. For this reason, it 
has been extensively applied to a variety of aspects concerning 
the dynamics of mesons and baryons made of both heavy and light quarks.
  
Due to the flavor-spin symmetry connecting matrix elements for different 
heavy flavors and angular momenta, the leading order description is 
simple and intuitive, and involves a few nonperturbative quantities in terms 
of which a large number of processes can be described. This leading 
order seems to be consistent with the  main features of heavy hadron 
physics as they are  presently observed. 
However, although being suppressed by the large heavy quark mass, the 
next-to-leading order corrections  proportional to $1/m_Q$ are also expected 
to play a significant role. In the framework of HQET, they 
are limited in number, and can be systematically identified and related 
to nonperturbative parameters having a clear physical meaning. 
A noticeable example can be found in 
the effective QCD heavy quark Lagrangian which,  up to order 
$1/m_Q$, can be written as 
\begin{equation}
{\cal L}~=~{\bar h_v} i (v \cdot D) h_v + \frac{1}{2 m_Q} {\cal K}
+ \frac{1}{2 m_Q} {\cal S}
+O(\frac{1}{m_Q^2}). \label{lag}
\end{equation} 
Here $h_v$ is the effective field for a heavy quark with velocity $v$, 
related to the Dirac field $Q$ by 
\begin{equation}
{ h_v}(x)~=~e^{i m_Q v \cdot x} \frac{1+ {\rlap v}/}{2} Q(x) \label{hv},
\end{equation}
${\cal K}$ is the nonrelativistic kinetic energy operator defined by 
\begin{equation}
{\cal K}~=~{\bar h_v} (i D^\perp)^2 h_v \label{kin}
\end{equation}
where $(D^\perp)^2=D_\mu D^\mu -(v\cdot D)^2$, with $D^\mu$ the QCD covariant 
derivative, and ${\cal S}$ is the Pauli 
term responsible of the spin splitting among the states belonging to the same 
heavy hadron multiplet in the $m_Q \to \infty$ limit: 
\begin{equation} 
{\cal S}~=~\frac{1}{2}
\left(\frac{\alpha_s\left(m_Q\right)}
{\alpha_s\left(\mu\right)}\right)^{3/\beta_0}
~{\bar h_v} \sigma^{\mu\nu} g G_{\mu\nu} h_v. \label{cromo}
\end{equation}
Appropriately normalized matrix elements of the operators (\ref{kin}) and 
(\ref{cromo}) account for the next-to-leading corrections  to the heavy 
hadron mass spectra: 
\begin{equation}
m_H~=~m_Q + \Delta - \frac{1}{2m_Q}\langle{\cal K}\rangle_H - 
\frac{c_H}{2m_Q}\langle{\cal S}\rangle_H, \label{mass}
\end{equation}
where $\Delta$ represents the binding energy of light quarks in the static 
approximation and $c_H$ is a numerical coefficient.  At the same time, 
they are involved as `preasymptotic' corrections in the $1/m_Q$ expansion of 
inclusive heavy hadron decay rates based on parton model ideas and on the 
operator product expansion \cite{shifman,bigi}. For example, the inclusive
$H \to f$ weak decay rate can be written as:  
\begin{equation} 
\Gamma(H\to f)~=~\frac{G_F^2m_Q^5}{192\pi^3}\vert V_{CKM}\vert^2
\left[c_3^f\left(1+\frac{\langle{\cal K}\rangle_H}  
{2m_Q^2}\right)+c_5^f\frac{\langle{\cal S}\rangle_H}{2m_Q^2} + 
O(\frac{1}{m_Q^3})\right],
\label{rate}
\end{equation}
where $c_3^f$ and $c_5^f$ are calculable short-distance coefficients
that depend 
on the process under consideration.\par 
The matrix elements of $\cal K$ and $\cal S$ are nonperturbative 
parameters that should be either determined phenomenologically from 
experimental data 
or estimated using a nonperturbative theoretical approach. In the case 
of heavy mesons ($P_Q$), information on both 
$\langle{\cal K}\rangle_{P_Q}$ and
$\langle{\cal S}\rangle_{P_Q}$ is available. Indeed, the 
chromomagnetic operator matrix element can be determined from the 
vector-pseudoscalar meson mass difference, indicating 
\begin{equation}
\langle{\cal S}\rangle_{P_Q} = \frac{3}{4}\left(m^2_{V_Q}-m^2_{P_Q}\right)
\simeq 0.37\hskip 2pt GeV^2, \label{masdif}
\end{equation}
and it has also been theoretically estimated by QCD sum rules giving a result 
in agreement with this value \cite{ball}. Concerning the kinetic energy 
operator matrix 
element, general arguments imply the inequality \cite{bigi1}
$\vert\langle{\cal K}\rangle_{P_Q}\vert\ge\langle{\cal S}\rangle_{P_Q}$ giving 
$\vert\langle{\cal K}\rangle_{P_Q}\vert\ge 0.37\hskip 2pt GeV^2$ using 
(\ref{masdif}), and the most recent 
theoretical estimate using QCD sum rules indicates \cite{ball}
\footnote{For earlier estimates see \cite{old}.}
\begin{equation} 
\langle{\cal K}\rangle_{P_Q} = - (0.60 \pm 0.10) \hskip 2pt GeV^2 \; .
\label{meskin}
\end{equation}
Finally, the first calculation with lattice QCD gives the inequality 
$|\langle{\cal K}\rangle_{P_Q}|\hskip 2pt <\hskip 2pt 1\hskip 2pt \; GeV^2$, in 
agreement with the above numbers \cite{marti}.\par 
On the other hand, no analogous information is available for the baryonic 
matrix elements of ${\cal S}$ and ${\cal K}$. In fact, the heavy baryon mass 
splittings relevant to the determination of the chromomagnetic expectation 
value are not measured yet; moreover, general lower bounds on 
the kinetic energy have not been worked out in this case. 
To leading order in $1/m_Q$, 
the heavy quark kinetic energy inside the baryon can be connected to that in 
the meson and to the heavy quark masses by the relation (which assumes that the 
charm mass $m_c$ is heavy enough for the expansion up to 
$1/m_c$ to be meaningful) \cite{bigi2}: 
\begin{equation} 
\langle{\cal K}\rangle_{P_Q}-\langle{\cal K}\rangle_{\Lambda_Q}\simeq 
\frac{m_b m_c}{2 (m_b-m_c)}\left[\left(m_B+3m_{B^*}-4m_{\Lambda_b}\right)-
\left(m_D+3m_{D^*}-4m_{\Lambda_c}\right)\right].\label{diff}
\end{equation}
Using present data, Eq.~(\ref{diff}) would give 
$\langle{\cal K}\rangle_{P_Q}-\langle{\cal K}\rangle_{\Lambda_Q}\simeq
0.07\pm 0.20~GeV^2$, where the error mainly comes from the error on the
the measured $m_{\Lambda_b}$, that  is not 
small enough:
$m_{\Lambda_b} = 5641 \pm 50 \; MeV$ \cite{PDG} .
Arguments have been given in favour of the near 
equality of  the kinetic energy operator matrix elements
between meson and baryon states \cite{manohar}, which could 
also be expected on intuitive grounds, but 
some quantitative 
estimates are needed. Here, we would like to consider in this regard 
the simplest case, represented by the $\Lambda_b$ baryon, where the 
chromomagnetic term vanishes in the infinite heavy quark mass limit
$\langle{\cal S}\rangle_{\Lambda_b} = 0$ as the light 
diquark system carries no spin (the same is true for the $\Xi_Q$), and 
attempt an assessment of the heavy quark kinetic energy. Specifically, we 
discuss the possibility of such an estimate in the framework of QCD sum rules, 
where the calculation has also been done for the meson, so that it is 
of interest to compare the two cases.

\section{Heavy baryon current and two-point function}
A basic element in the application of QCD sum rules to baryons is the 
definition of a suitable baryon interpolating effective field in terms of 
quark fields, and the corresponding vacuum-to-baryon matrix element. For the 
case of heavy baryons this problem has been discussed previously in 
Refs.~\cite{shuryak}-~\cite{bagan}, and we outline here this analysis for the 
sake of completeness.
 
The general expression of the interpolating field for a baryon state 
with one heavy quark and two light quarks can be written as 
\begin{equation}
J^{ (\ell) }_D~=~\epsilon^{\alpha \beta \gamma} (q_1^{T \alpha } C 
\Gamma^{ (\ell) } \tau q_2^\beta) {\tilde \Gamma} (h_v^\gamma)_D
\end{equation}
where $T$ means transpose,
$\alpha,\hskip 2pt \beta$ and $\gamma$ are color indices, $D$ is a 
Dirac index, $C$ is the charge conjugation matrix, $q_1$ and  $q_2$ are 
light quark 
field operators, and $\tau$ is a flavor matrix. For the case of the spin 
$1/2$ ${\Lambda_b}$ of interest here, the matrices $\tau$ and ${\tilde \Gamma}$ are 
given by:
\begin{equation} 
{\displaystyle {\tau_{i j}= \frac{1}{\sqrt 2}(\delta_{1 i} \delta_{2 j}-
\delta_{2 i} \delta_{1 j})}}, \qquad {\tilde \Gamma} =1,\label{tau}
\end{equation}
whereas two choices are possible for the matrix $\Gamma^{(\ell)}$: 
\begin{equation} 
\Gamma^{(1)}=\gamma_5, \qquad \Gamma^{(2)}=\gamma_5 \gamma_0.\label{gamma}
\end{equation} 
Such choices correspond to spin-parity $J^P=0^+$ for the light quark pair, 
as implied by the predicted spectroscopy of baryons containing one heavy 
quark in the infinite mass limit \cite{yan,georgi}.\par 
Consequently, the general ${\Lambda_b}$ interpolating field is the linear 
combination 
\begin{equation}
{J}={J}^{(1)}+ b \; \; {J}^{(2)},\label{j}
\end{equation}
depending on the, a priori unspecified, parameter $b$. 

Together with the baryonic interpolating currents, we need for our estimates
 the values of the vacuum-to-baryon couplings  (${\ell}= 1,2$):
\begin{equation}
<0 | J^{(\ell)}_D | {\Lambda_b} (v) > = f_{{\Lambda_b} }^{(\ell)} (h_v)_D. 
\label{sda}
\end{equation}
In the framework of QCD sum rules these constants can be obtained from the 
two-point correlators: 
\begin{equation}  
\Pi_{CD}^{(\ell, \ell^\prime)}(k) 
= i \int dx <0| T(J^{(\ell)}_{C}(0){\bar J}^{(\ell^\prime)}_{D}(x) 
|0> e^{i k x}
\label{corr}
\end{equation}
where, according to the familiar rules of the heavy quark theory, $k$ is the 
residual momentum, corresponding to the decomposition $p^\mu=m_b v^\mu+k^\mu$ 
of the heavy baryon momentum; we parametrize it as
\begin{equation}
k^\mu = \omega ~v^\mu. \label{k}
\end{equation}
Taking Eqs.~(\ref{hv}) and (\ref{k}) into account, we can write the LHS of 
Eq.~(\ref{corr}) as
\begin{equation}
\Pi_{CD}^{(\ell, \ell^\prime)}(k)= (1+ {\rlap v}/)_{CD}\Pi^{(\ell, \ell^\prime)}
(\omega). \label{cor}
\end{equation} 
The scalar functions $\Pi^{(\ell, \ell^\prime)}$ satisfy  dispersion 
relations of the form:
\begin{equation}
\Pi^{(\ell, \ell^\prime)} (\omega) =\frac{1}{\pi} \int d \Omega ~
\frac{\rho^{(\ell, \ell^\prime)} (\Omega )}{ \Omega - \omega }
~+~subtractions, \label{dr}
\end{equation}
where $\rho$ represents the spectral function:  
$\rho^{(\ell, \ell^\prime)} (\Omega) = Im \Pi^{(\ell, \ell^\prime)} 
(\Omega)$. The explicit form of the polynomial representing the subtractions 
is not important because in the sequel we shall apply the borelized version of 
the sum rule, and the Borel transform of any polynomial vanishes.\par 
Following the QCD sum rules method \cite{shifman1}, 
$\rho^{(\ell, \ell^\prime)} (\Omega)$
can be computed in two ways: either by saturating the dispersion
relation (\ref{dr}) by physical hadronic states or by means of the operator 
product expansion (OPE), which is allowed insofar $k^2$ in (\ref{corr}) is
large and negative.\par 
Performing the OPE on $\Pi^{(\ell, \ell^\prime)}$ and taking into account
the operators of dimension $d=0$  (the perturbative term), 
$d=3$ and $d=5$, by a straightforward calculation in the heavy quark theory 
one finds the spectral functions:
\begin{equation}
\rho^{(\ell, \ell)}(\Omega) =  \rho^{(0)}(\Omega)  =
\frac{\Omega^5}{20 \pi^3} \label{cntr1}
\end{equation}
for $\ell = \ell^\prime$ (`diagonal' sum rules), and 
\begin{equation}
\rho^{(1, 2)}(\Omega) + \rho^{(2, 1 )}(\Omega) = 
\rho^{(3)}(\Omega)+ \rho^{(5)}(\Omega) = 
-\frac{2\langle{\bar q}q\rangle\Omega^2}{\pi} -
\frac{2 m_0^2\langle{\bar q}q\rangle}{\pi}
\left(-\frac{1}{8} + \frac{1}{16}\right) 
\label{cntr2}
\end{equation}
for  $\ell\neq\ell^\prime$ (`nondiagonal' sum rules),
as obtained previously \cite{grozin,grozin1,bagan}. In Eqs.~(\ref{cntr1}) and 
(\ref{cntr2}) $\rho^{(0,3,5)}$ denote, respectively,  the leading 
perturbative term in the short distance expansion and the $d=3$ and $d=5$ 
contributions. Moreover, $\langle{\bar q}q\rangle$ is the quark condensate 
($\langle{\bar q}q\rangle =-(230\hskip 2pt MeV)^3$ at the scale of 
$1\hskip 2pt GeV$) and $m_0$ is connected to the quark-gluon condensate 
by the relation $\langle{\bar q}g_s\sigma \cdot G q\rangle =m_0^2 
\langle{\bar q}q\rangle$ (and $m_0^2=0.8\hskip 2pt GeV^2$ \cite{shifman1}).  
The separation of the $d=5$ contribution in two contributions reflects the 
existence of two independent sources originating such a term, namely the 
nonlocal quark condensate \cite{rad}
\begin{equation}
\langle{\bar q}(x)q(0)\rangle=\Phi(x^2)\langle{\bar q}q\rangle \simeq
\left(1+\frac{x^2 m_0^2}{16}\right) \langle{\bar q}q\rangle
\label{local}
\end{equation}
and the diagram where the gluon non perturbative field is emitted from 
the perturbative light quark line (in the adopted Fock-Schwinger gauge,  
no such gluon can be emitted from the heavy quark leg in the infinite heavy 
quark mass limit).
 
Actually, in previous calculations also the $d=4$ gluon condensate 
$\langle\alpha_s G^2\rangle$ and the higher dimension operators with $d>5$ 
have been included. 
The possibility of potentially non-negligible higher order terms is briefly
discussed at the end of this Section.
On the other hand, the gluon condensate contribution would 
be rather easy to include both in Eq.~(\ref{cntr1}) and in the analogous 
three-point functions considered in the next Section to evaluate the kinetic 
energy. In principle, its presence accounts for the low momentum part of the 
gluon exchange diagram of order $\alpha_s$. 
In the baryon case the high momentum component, i.e. 
the $\alpha_s$ correction to the perturbative diagram, requires a three-loop 
calculation, which has not been attempted yet. Consequently, 
we do not include the gluon 
condensate as it would only partially represent the gluon exchange, 
and we limit to the analysis of both two-point and three-point correlators with 
the operators having $d=0,\hskip 2pt 3$ and $5$.

Turning to the representation of spectral functions in (\ref{dr}) in terms of 
physical hadronic states, we write 
\begin{equation}
\rho_{had}^{(\ell, \ell^\prime)} (\Omega)= \rho_{res}^{(\ell, 
\ell^\prime)} 
(\Omega)+ \rho_{cont}^{(\ell, \ell^\prime) } (\Omega),
\label{had}
\end{equation}
where $\rho_{res}^{(\ell, \ell^\prime)}(\Omega)$ contains the contribution
of the low-lying baryon state, in this case the ${\Lambda_b}$, and
$ \rho_{cont}^{(\ell, \ell^\prime)}$ represents the contribution of 
higher mass states. Assuming, as usual, duality between partons and 
resonances, $\rho_{cont}^{(\ell, \ell^\prime)}(\Omega)$ can be taken 
as equal to 
the operator product expansion $\rho^{(\ell, \ell^\prime)}(\Omega)$ of 
Eqs.~(\ref{cntr1}) and (\ref{cntr2}) for $\Omega >\omega_c$, with $\omega_c$ a 
continuum threshold, and vanishing elsewhere. As for 
$\rho_{res}^{(\ell, \ell^\prime)} (\Omega)$, they are expressed as: 
\begin{eqnarray}
\rho_{res}^{(\ell,\ell)} (\Omega)& =& \frac{\pi [f_{\Lambda_b}^{ (\ell ) } ]^2}{2}
\delta(\Omega-\Delta)\\
\rho_{res}^{(1,2)} (\Omega) + \rho_{res}^{(2,1)} (\Omega) & =& 
\frac{\pi f_{\Lambda_b}^{(1)} f_{\Lambda_b}^{(2)}}{2}
\delta(\Omega-\Delta)
\end{eqnarray}
where $f_{\Lambda_b}^{(\ell)}$ have been defined in Eq.~(\ref{sda}) and 
$\Delta= m_{{\Lambda_b}} - m_b$ in the limit $m_b \to \infty$ (see Eq.~(\ref{mass})). 
Equating the two representations of the two-point correlators, and taking the 
Borel transform in the variable $\omega$ of both sides, one obtains the set of sum 
rules for the vacuum-to-baryon couplings of the interpolating fields:
\begin{eqnarray}
[ f^{(1)}_{\Lambda_b} ]^2  &=& [ f^{(2)}_{\Lambda_b} ]^2 =
\frac{2}{\pi} \int d \Omega~e^{-\frac{\Omega - \Delta}{E}} 
\rho^{(0)}(\Omega)  \label{twopd}\\
f^{(1)}_{\Lambda_b}  f^{(2)}_{\Lambda_b} &=& 
\frac{1}{\pi} \int d \Omega~e^{-\frac{\Omega - \Delta}{E}} 
[\rho^{(3)}(\Omega)+ \rho^{(5)}(\Omega)]. 
\label{twopnd}
\end{eqnarray}
One can notice that, due to the structure of Eq.~(\ref{cntr1}), at the chosen 
order of approximation in the operator product expansion 
$\vert f_{\Lambda_b}^{(1)}\vert\simeq\vert f_{\Lambda_b}^{(2)}\vert =
f_{\Lambda_b}$ from the `diagonal' 
sum rules.\footnote{As it will be seen in the next Section, this does no longer 
occur in the case of the three-point function, in the sense that somehow
different results for the kinetic energy are obtained from the two
diagonal sum rules.} 
This result should be 
confirmed by the non-diagonal sum rule, which can also be used to fix the 
relative sign between the two constants. \par 
In order to extract useful information from the sum rules, we have to fix 
the parameters $\Delta,\hskip 2pt \omega_c$ and the allowed range for the 
Borel parameter $E$ (the so-called `duality' region). Indeed, in order to 
obtain a prediction for $\Delta$, we differentiate (\ref{twopd})
( or (\ref{twopnd}) ) 
in the variable 
$1/E$ and take the ratio with (\ref{twopd}) (or (\ref{twopnd}) ) itself, 
thus obtaining a new sum rule where only $\Delta$ (and not $f_{\Lambda_b}$) 
appears.
\par  
The results for $f_{\Lambda_b}$ and $\Delta$ as a function of the Borel 
variable $E$, 
and for $\omega_c$ in the range $1.1 - 1.3\hskip 2pt GeV$, are shown in Fig.~1. 
In particular, in Fig.~1a we depict the result for the diagonal correlators
(\ref{twopd}) both concerning the coupling $f_{\Lambda_b}$ and $\Delta$, 
in Fig.~1b the outcome of the nondiagonal correlator (\ref{twopnd}).
In Fig.~1c we display 
the result from the sum rule for the correlator of the current 
$J$ in Eq.(\ref{j}):
\begin{equation}
{1\over 2} f^2_{\Lambda_b} (1+b)^2 e^{-\Delta/E} =
\int_0^{\omega_c} d \omega \; e^{-\omega/E} 
\Big [  {1 + b^2 \over 20 \pi^4} \omega^5 - 
{2 b <{\overline q}q> \over \pi^2}  
( \omega^2 - { m_0^2 \over 16}) \Big] \label{twop2}
\end{equation}
with the choice $b=1$, that we shall justify in the next Section.

As one can see, stability is always reached for values of $E$ from 
$0.3\hskip 2pt GeV$ on, which at first sight would be a welcomed property. 
However, the familiar criterion that the continuum should not be too large 
with respect to the ground state baryon contribution, favoring low values of 
$E$, rather strongly reduces the `working' region to a range in 
the neighborhood of $E=0.4 - 0.5 \; GeV$. 
In the next Section
this effect of the continuum will be found to be even more significant
in the kinetic energy calculation using  
three-point sum rules. Clearly, it originates from the 
high power dependence of the  
spectral functions, that is expected on dimensional grounds.
In this regard, considering 
the structure of the operator product expansion in 
Eqs.~(\ref{cntr1}) and (\ref{cntr2}),
it should be noticed  that the suppression of dimensional 
condensates is provided by smaller powers of the integration variable 
$\Omega$, compared to the leading perturbative term, rather than by large 
mass denominators characterizing the applications of the QCD sum rule method 
with finite heavy quark mass. 

Coming to the numerical results, it is worth observing that the diagonal and 
non-diagonal correlators, together with
 the correlator of the current $J$ in (\ref{j}),
provide us with the same mass parameter $\Delta_{\Lambda_b}$:
\begin{equation}
 \Delta_{\Lambda_b} = 0.9 \pm 0.1 \; GeV \; .\label{delta} \end{equation}
The uncertainty on $\Delta_{\Lambda_b}$ in 
(\ref{delta}) comes from the variation of the continuum threshold and
from changing the Borel parameter $E$ in the range $E=0.3 - 0.4 \; GeV$
in Fig.1a and c,  and $E=0.3 - 0.5 \; GeV$ in Fig.1b. 
Using the result of the analysis for the mesonic system
$ \Delta_B \simeq 0.5\; GeV $ \cite{ball} we get for ${\Lambda_b}$ the mass
$M_{\Lambda_b} = M_B + \Delta_{\Lambda_b} - \Delta_B \simeq 5.7 \; GeV$, 
to be compared with the experimental measurement quoted above.
The difference of
$200 \; MeV$ between the mass of the ${\Lambda_b}$ and the continuum 
threshold is of reasonable size as it
nearly corresponds to the mass of a pair of pions.

As for the coupling $f_{\Lambda_b}$, the diagonal and non-diagonal sum rules in
(\ref{twopd}) and (\ref{twopnd}) give 
rather different results: 
\begin{eqnarray}
f_{\Lambda_b} &=& (2.0 \pm 0.5) \times 10^{-2} \; GeV^3 \;\;\;  \\ 
f_{\Lambda_b} &=& (3.5 \pm 0.5) \times 10^{-2} \; GeV^3 \;\;\; \; \; \label{fl}
\end{eqnarray}
respectively.
The different results can be put in better agreement using a smaller value for 
the quark condensate, as advocated in \cite{grozin}. However,
from the correlator of the current $J$, with $b=1$,  we obtain:
\begin{equation}
 f_{\Lambda_b} = (2.9 \pm 0.5) \times 10^{-2} \; GeV^3 \; ,
\label{29} \end{equation}
that is the result we shall use in the next Section.

As far as higher dimensional contributions to (2.20)-(\ref{29}) are concerned, 
it 
was observed in \cite{grozin,bagan} that the most important one 
should be the four-quark $d=6$ operator contributing to the diagonal sum rule 
(\ref{cntr1}) and to the sum rule (\ref{twop2}). 
The evaluation of the relevant diagram
 is straightforward and leads to a term proportional to
$<{\bar q}(x) q(0) {\bar q}(x) q(0)>$. The numerical analysis can be done 
by making
the assumption of factorization of the nonlocal four-quark
condensate and, using this hypothesis, 
we would obtain a correction to (2.20) and (\ref{29}) whithin the quoted 
uncertainty.
However, the accuracy of this assumption is difficult to assess
due to the fact that the numerical result would strongly depend on the 
explicit model for the nonlocal condensate
and, as a matter of fact, the inclusion of this term 
has been questioned, e.g. in ref.\cite{bagan}. For this reason, following 
\cite{bagan},  we prefer to
omit this term, with the understanding that this neglect will be reflected
in an additional uncertainty in the final result.

\section{The three-point function and $\langle{\cal K}\rangle_{\Lambda_b}$}
To apply the formalism of QCD sum rules to the kinetic energy operator 
${\cal K}$, we consider the three-point correlators:
\begin{equation}
\Sigma_{CD}^{(\ell, \ell^\prime)}(k, k^\prime ) =(1+ {\rlap v}/)_{CD} 
\Sigma^{(\ell, \ell^\prime)}(\omega, \omega^\prime)
= i^2 \int dx dy <0| T(J^{(\ell)}_{C}(y) {\cal K}(0)
{\bar J}^{(\ell^\prime)}_{D}(x) 
|0> e^{-i k y +i k^\prime x},
\label{corr2}
\end{equation}
where $(l,\hskip 1ptl^\prime)=1,\hskip 1pt 2$ have the same meaning as in 
Eq.~(\ref{corr}), and $k,\hskip 1pt k^\prime$ are the residual momenta, 
corresponding to the decomposition of the heavy baryon momentum
$p^\mu=m_b v^\mu +k^\mu$, 
$p^{\prime \mu}=m_b v^\mu +k^{\prime \mu}$.  
Analogously to Eq.~(\ref{k}), we have taken:  
\begin{equation}
k^\mu = \omega~ v^\mu, ~~~~k^{\prime \mu} = \omega^\prime~ v^\mu.
\end{equation}
The scalar function $\Sigma^{(\ell, \ell^\prime)}$ satisfies a  double 
dispersion relation in the variables $\omega,\hskip 1pt \omega^\prime$:
\begin{equation}
\Sigma^{(\ell, \ell^\prime)}(\omega, \omega^\prime )
 =\int d \Omega ~
d \Omega^\prime ~
\frac{\sigma^{(\ell, \ell^\prime)} (\Omega, \Omega^\prime )}
{( \Omega - \omega)( \Omega^\prime - \omega^\prime) }
~+~subtractions \label{disp}
\end{equation}
where, as for the two-point function, the explicit form of the subtraction term 
is not important since it will be eliminated by the double Borel transform of 
the sum rule.\par 
For both $k^2$ and $k^{\prime 2}$ large and negative, so that short distances 
$x$ and $y$ should dominate, we perform the operator product expansion on the 
RHS of Eq.~(\ref{corr2}) taking into account only the operators of dimension 
$d=0$ (the perturbative term), $d=3$ and $d=5$, analogously to Sec. 2, and 
neglecting the $d=4$ gluon condensate with the same motivation given there. 
In this approximation we obtain for the various spectral functions the 
following expressions, applying the Cutkosky rules in momentum space:
\begin{eqnarray}
\sigma^{(1,1)}(\Omega, \Omega^\prime) &=& -\frac{3 \Omega^7}{140 \pi^4}
\delta (\Omega - \Omega^\prime)       \\
\sigma^{(2,2)}(\Omega, \Omega^\prime) &=& -\frac{5 \Omega^7}{140 \pi^4}
\delta (\Omega - \Omega^\prime)    \\
\sigma^{(1,2)} (\Omega) +
\sigma^{(2,1)}(\Omega) &=& \left[ \frac{2 <{\overline q}q>}{\pi^2} \Omega^4 
+ \frac{2 m_0^2 <{\overline q}q>}{\pi^2} \Omega^2 \left(-\frac{3}{8} + 
\frac{3}{16}\right) \right]\delta (\Omega-\Omega^\prime).\label{c1}
\end{eqnarray}
Thus, $\sigma^{(1,1)}$ and $ \sigma^{(2,2)}$ only contain the 
leading perturbative term in the short distance expansion, and  $d=3$ and 
$d=5$ condensates do not contribute to the operator product expansion of 
these ``diagonal' correlators. 
Also the $d=6$ four-quark operator does not contribute in this case.
On the other hand, the `nondiagonal' spectral 
function $\sigma^{(1,2)} (\Omega) + \sigma^{(2,1)} (\Omega)$ is determined by  
the $d=3$ $\langle {\bar q} q\rangle$ vacuum condensate and by the $d=5$ 
quark-gluon condensate. Similar to Eq.~(\ref{cntr2}), for the latter 
condensate the two separate contributions in (3.6) arise from the non local 
contribution of Eq.~(\ref{local}) and from the diagram where the non 
perturbative low-frequency gluon is emitted from the perturbatively 
propagating light quark.\par 
The hadronic side of the sum rule is written as
\begin{equation}
\sigma_{had}^{(\ell, \ell^\prime)} (\Omega, \Omega^\prime)= 
\sigma_{res}^{(\ell, 
\ell^\prime)} 
(\Omega, \Omega^\prime)+ \sigma_{cont}^{(\ell, \ell^\prime) } (\Omega, 
\Omega^\prime)~,
\label{had2}
\end{equation}
where $\sigma_{res}^{(\ell, \ell^\prime)} (\Omega, \Omega^\prime)$ 
contains the contribution of the ${\Lambda_b}$ ground state and
$ \sigma_{cont}^{(\ell, \ell^\prime)}$ contains the contribution of
higher mass states. By applying the duality principle, 
$ \sigma_{cont}^{(\ell, \ell^\prime)}$ is assumed equal to the operator 
product expansion $\sigma^{(\ell, \ell^\prime)}(\Omega, \Omega^\prime)$ 
of Eqs.~(\ref{c1})-(3.6) in the domain $\Omega >\omega_c$  and 
$\Omega^\prime >\omega_c$, and vanishing outside this domain. Owing to the 
$\delta$-function behavior, the details of the integration domain far from 
the diagonal $\Omega=\Omega^\prime$ should not be important.\par 
The pole contribution is as follows:
\begin{eqnarray}
\sigma_{res}^{(\ell, \ell)} (\Omega, \Omega^\prime) &= &
\frac{ [ f_{\Lambda_b}^{(\ell)} ]^2 {\langle{\cal K}\rangle}_{\Lambda_b}}
{2 (\Omega - \Delta) (\Omega^\prime - \Delta)},\\
\sigma_{res}^{(1, 2)} (\Omega, \Omega^\prime) +
\sigma_{res}^{(2, 1)} (\Omega, \Omega^\prime) &= &
\frac{ f_{\Lambda_b}^{(1)} f_{\Lambda_b}^{(2)} 
{\langle{\cal K}\rangle}_{\Lambda_b}}{2 (\Omega - \Delta)
(\Omega^\prime - \Delta)}, 
\end{eqnarray}
where the matrix element $\langle{\cal K}\rangle_{\Lambda_b}$ is defined as:
\begin{equation}
{\langle{\cal K}\rangle}_{\Lambda_b} = \langle \Lambda_b (v) 
|{\cal K}| \Lambda_b (v) \rangle =
\langle \Lambda_b (v) |{\bar h_v} (i D^\perp)^2 h_v | \Lambda_b (v) \rangle \; .
\label{KK}
\end{equation}
\par 
Sum rules for $\langle{\cal K}\rangle_{\Lambda_b}$ are obtained by taking 
double Borel transforms 
in the variables $\Omega, \Omega^\prime$ of both $\Sigma_{OPE}$ 
and $\Sigma_{res}$, giving : 
\begin{equation} 
{\cal B}(E)\frac{1}{\omega-\Omega}=\frac{1}{E} e^{-\omega/E},\qquad
{\cal B}(E)\frac{1}{\Delta-\Omega}=\frac{1}{E} e^{-\Delta/E}, 
\label{borel}
\end{equation}
and similar for $\Omega^\prime$, and by assuming that one can equate them for 
some range of the Borel parameters $E_1$, $E_2$. The symmetry of the spectral 
functions in $\Omega$, $\Omega^\prime$ suggests the choice $E_1=E_2$ and, 
moreover, we take $E_1=E_2=2E$ with $E$ the Borel parameter of the sum 
rule for the two-point correlator, by analogy with the calculation of the 
Isgur-Wise form factor, both for baryons \cite{grozin1}
and mesons \cite{neubert1},  where the normalization at zero 
recoil requires such a relation. Finally, we assume the same continuum 
threshold $\omega_c$ as in the two-point function sum rule, 
namely $\omega_c=1.1,  \; 1.2$ and $1.3 \; GeV^2$.
  
\par 
In this way, eliminating the couplings $f_{\Lambda_b}^{(\ell)}$ by means of the 
two-point sum rules in  Eqs.~(2.16) and (2.17), we obtain for the particular 
choices of the baryonic interpolating field the following sum rules for the 
matrix element ${\langle{\cal K}\rangle}_{\Lambda_b}$:
\begin{eqnarray}
{\langle{\cal K}\rangle_{\Lambda_b}}&=&-\frac{3}{7}~ \frac {I_7 } {I_5}
~~~~~~~~~~ ~~~~~~~~~~ ~~~~~~~~~~ (\ell= \ell^\prime=1) \label{trep1}\\
{\langle{\cal K}\rangle_{\Lambda_b}}&=&-\frac{5}{7}~ \frac {I_7 } {I_5}
~~~~~~~~~~ ~~~~~~~~~~ ~~~~~~~~~~ (\ell= \ell^\prime=2) \label{trep2}\\
{\langle{\cal K}\rangle_{\Lambda_b}}&=&     -\frac{I_4 ~ - ~ \frac{3}{16}~ 
m_0^2 ~ I_2}
{I_2 ~ - ~ \frac{1}{16}~ m_0^2 ~ I_0}~~~~~~~~~~ ~~~~~~~~~~ ~~ (\ell
\neq \ell^\prime) \label{trep3}
\end{eqnarray}
where
\begin{equation}
I_n~=~ \int_0^{\omega_c} d \omega \; \omega^n \; e^{-\frac{\omega}{E}} \; \; .
\end{equation}
\label{k2}
\noindent
In these relations, the dependence on $\Delta$ has disappeared, and in 
principle the analysis could now follow the familiar procedure, 
checking the existence of a range in the Borel parameter $E$ 
where the results for ${\langle{\cal K}\rangle_{\Lambda_b}}$ 
are stable or at least smoothly dependent on 
$E$, and at the same time the operator product expansion displays a 
convergent structure with higher states giving smaller contributions.  
However, in the present case the two `diagonal' sum rules 
(\ref{trep1}) and (\ref{trep2}) 
do not give the same result for ${\langle{\cal K}\rangle_{\Lambda_b}}$, 
at least in the 
approximation of the operator product expansion considered here, in contrast 
with the two-point function sum rules in (\ref{twopd}). 
Thus, the separation of 
`diagonal' and `nondiagonal' sum rules is no longer useful, and we have to 
use the complete three-point correlator with the full $\Lambda_b$ 
interpolating field of Eq.~(\ref{j}), where the {\it a priori} unknown 
parameter $b$ appears. Clearly, the corresponding sum rule 
\begin{equation}
{1\over 2} {\langle{\cal K}\rangle}_{\Lambda_b} f^2_{\Lambda_b} (1+b)^2 
e^{-\Delta/E} =
\int_0^{\omega_c} d \omega e^{-\omega/E} 
\Big [ - {3 + 5 b^2 \over 140 \pi^4} \omega^7 + 
{2 b <{\overline q}q> \over \pi^2} \omega^2 ( \omega^2 - {3 m_0^2 \over 16}) 
\Big]
\label{trep}
\end{equation}
would give results 
depending on the parameter $b$, whose value must be determined somehow.
\par 
In order to get a reasonable criterion to fix the value of $b$, we recall 
that, in principle, the three-point function sum rule can also be used to 
determine the parameter $\Delta$, by taking the derivative in $1/E$ of 
(\ref{trep}). 
We can then compare such determination with that given by the 
two-point function sum rule (\ref{twop2})
and assume, as an `optimal choice', the value of 
$b$ for which the two determinations are closer. In Fig.~2 we depict the 
ratio of the two determinations ${\it versus}$ $b$ and the Borel parameter $E$. 
As one can see, for the considered values of $E$ such a ratio is close to 
unity for $b=1$, and therefore we shall assume this value of $b$ for 
the subsequent analysis 
of the sum rule for ${\langle{\cal K}\rangle_{\Lambda_b}}$.
\par 
In Fig.~3 we report the results of the analysis of the sum rule (\ref{trep})
for $b=1$, 
corresponding to the chosen three values of the continuum threshold 
$\omega_c$. 
In this analysis we use (\ref{twop2}) for $f_{\Lambda_b}$ thus avoiding the 
assumption $f_{\Lambda_b}^{(1)}=f_{\Lambda_b}^{(2)}$.
There is a remarkable plateau for $E$ larger than 
$0.25\hskip 2pt GeV$ indicating a value of 
${\langle{\cal K}\rangle_{\Lambda_b}}$ around 
$-0.6\hskip 2pt GeV^2$, with an uncertainty of $0.1 \; GeV^2$ coming from the 
variation of the continuum threshold in the range $1.1 - 1.3 \; GeV$.
However, to avoid upsetting the usual QCD sum rule 
criteria by an excessive contribution from the continuum, in practice 
the real `window' in $E$ is reduced to a small range around 
$E\sim 0.3\hskip 2pt GeV$. This feature, which is due
 to the highly divergent behavior in 
$\Omega$ of the spectral functions, was already anticipated by the analysis of 
two-point function sum rules. Clearly, 
such a low value of the Borel parameter would call for the calculation 
of the order $\alpha_s$ correction in both the three-point and in the 
two-point sum rules, a task that goes beyond the aim of the present paper. 
In the case of the beauty meson $B$, it is worth reminding that the radiative
corrections turn out to be 
individually important in two and three point functions, but they compensate
in the expression for $\langle{\cal K}\rangle$ confirming the indication of the 
lowest order \cite{ball}. 

In conclusion, within the approximation of the operator product expansion
adopted here, the result is
\begin{equation}
 {\langle{\cal K}\rangle}_{\Lambda_b} = - (0.6 \pm 0.1) \; GeV^2 \;. \label{1}
\end{equation}
Concerning the value quoted in (\ref{1}), we recall that the uncertainty there
only accounts for the effect of the variation 
of the continuum threshold and the Borel parameter; to it one should add the
theoretical uncertainty related to the neglect of the radiative corrections
and the $d=6$ operator in the two-point function.
We notice explicitly that the inclusion of the $d=6$ term, 
in the factorization hypothesis, could somehow enhance 
$f_{\Lambda_b}$ and reduce $|<{\cal K}>_{\Lambda_b}|$; however, this result 
would be numerically reasonable only for values of the Borel parameter $E$ 
larger than 
$E=0.4 \; GeV$, which would contradict our requirement on the 
continuum suppression. For these reasons, an additional 
uncertainty of the order of $\pm 0.1$ in (\ref{1}) cannot be excluded.

Let us finally observe that  a gratifying feature of the sum rules is
the consistency of the determinations of 
the mass parameter $\Delta$ from two and three-point functions.
We also notice that the result in (\ref{1})
supports the conjecture that the value of the heavy quark kinetic energy in the 
baryon is similar to that in the meson.

\acknowledgements

\noindent
We would like to thank P. Ball and V.M. Braun for interesting discussions.

\clearpage

\hskip 3 cm {\bf FIGURE CAPTIONS}
\vskip 1 cm

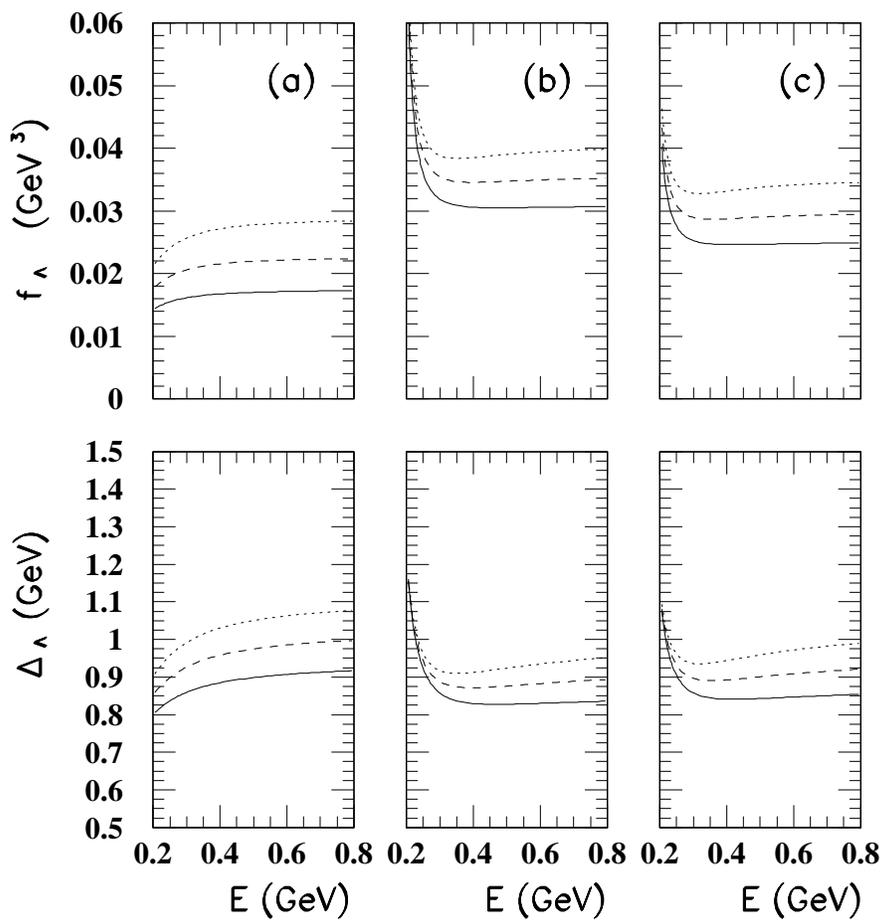
\noindent {\bf Fig. 1}\\
\noindent
Results from the two point sum rules Eq. (\ref{twopd}) (a),
Eq. (\ref{twopnd}) (b), and from the correlator of the current $J$
in Eq. (\ref{j}) with $b=1$ (c). The different curves 
correspond to different choices of the continuum threshold: 
$\omega_c= 1.1 \; GeV$ (continuous line),
$\omega_c= 1.2 \; GeV$ (dashed line),
$\omega_c= 1.3 \; GeV$ (dotted line). 
\vspace{5mm}

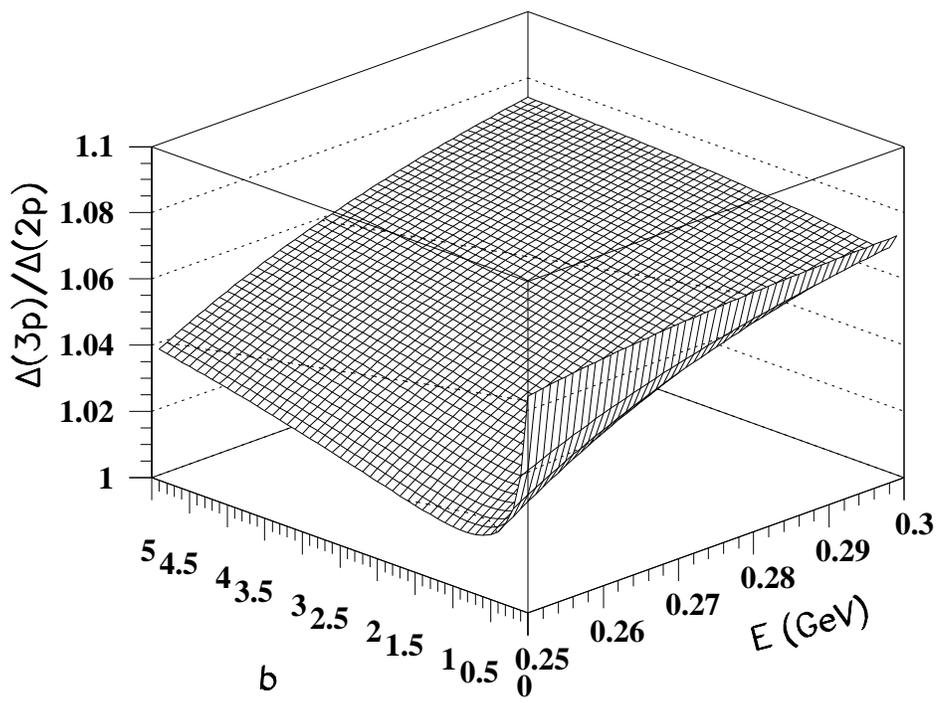
\noindent {\bf Fig. 2}\\
\noindent
The ratio of the mass parameters $\Delta$ obtained from three point (3p) 
and two points (2p) correlators of the current $J$
in Eq. (\ref{j}): ${\cal R}= \Delta(3p)/\Delta(2p)$. The continuum threshold 
is fixed to $\omega_c= 1.2 \; GeV$.
\vspace{5mm}
        
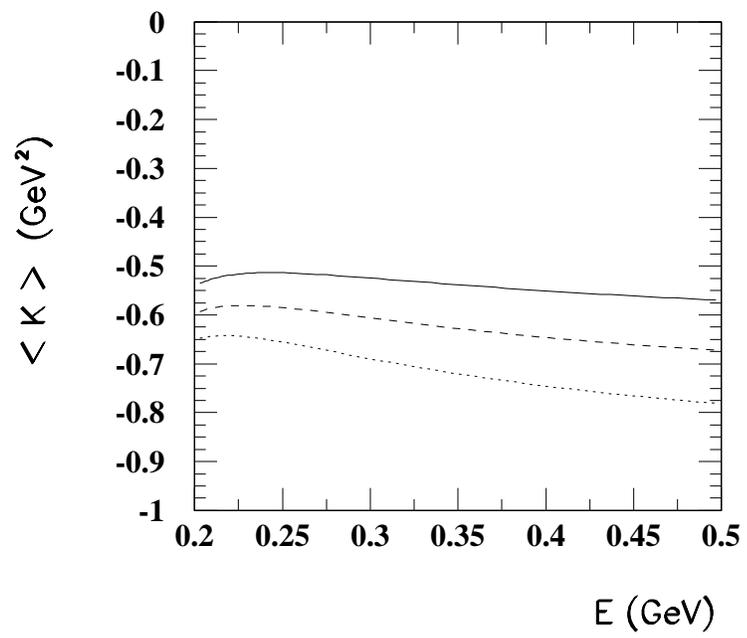
\noindent {\bf Fig. 3}\\
\noindent
$\langle{\cal K}\rangle_{\Lambda_b}$ 
from the three point correlator with the current $J$
in Eq. (\ref{j}) with $b=1$. The different curves correspond to the 
continuum thresholds: 
$\omega_c= 1.1 \; GeV$ (continuous line),
$\omega_c= 1.2 \; GeV$ (dashed line),
$\omega_c= 1.3 \; GeV$ (dotted line). 
\vspace{5mm}


\clearpage
\begin{references}
 
\bibitem{hqet}
For a review see:
M.B. Wise, in Proceedings of the 6th Lake Louise Winter Institute, Lake Louise, 
1991, edited by B.A. Campbell, A.N. Kamel, P. Kitching and F.C. Khanna
(World Scientific, Singapore, 1991), p. 222;
H. Georgi, in Proceedings of the TASI 1991, Boulder, 1991, edited by 
R.K. Ellis, C.T. Hill and J.D. Lykken (World Scientific, Singapore, 1992), 
p. 589.

\bibitem{shifman}
For a review and references to earlier analyses see:
B. Blok and M. Shifman, in Proceedings of the 3rd Workshop on the Tau-Charm 
Factory, Marbella, 1993, edited by J. Kirkby and R. Kirkby 
( Editions Frontieres,1994), p. 247.

\bibitem{bigi}
I.I. Bigi, N.G. Uraltsev, A. Vainshtein, Phys. Lett. B 293, 430 (1992); 
(E) B 297, 477 (1993);
B. Blok and M. Shifman, Nucl. Phys. B 399, 441 (1993); B 399, 459 (1993). 

\bibitem{ball}
P. Ball and V.M. Braun, Phys. Rev. D 49, 2472 (1994). 

\bibitem{bigi1} I.I. Bigi, M. A. Shifman, N.G. Uraltsev and A. Vainshtein, 
Phys. Rev. D 52, 196 (1995).

\bibitem{old} 
V. Eletski and E. Shuryak, Phys. Lett. B 276, 191 (1992);
M. Neubert, Phys. Rev. D 46 (1992) 1076; Phys. Lett. B 322, 419 (1994).

\bibitem{marti} M. Crisafulli, V. Gimenez, G. Martinelli and C.T. Sachrajda, 
report CERN-TH-7521-94, Rome 94/1071, SHEP 94/95-14.  


\bibitem{bigi2} I. Bigi, report UND-HEP-95-BIG02 (June 1995).

\bibitem{PDG} Review of Particle Properties, Phys. Rev. D 50 (1994).

\bibitem{manohar}  A.V. Manohar and M.B. Wise, Phys. Rev. D 49, 1310 (1994). 

\bibitem{shuryak} E.V. Shuryak, Nucl. Phys. B 198, 83 (1982).

\bibitem{grozin} A.G. Grozin and O.I. Yakovlev, Phys. Lett. B 285, 254 (1992).

\bibitem{grozin1} A.G. Grozin and O.I. Yakovlev, Phys. Lett. B 291, 441 (1992).

\bibitem{bagan} E. Bagan, M. Chabab, H.G. Dosch and S. Narison,
Phys. Lett. B 278, 367 (1992); B 287, 176 (1992); B 301, 243 (1993).

\bibitem{yan}
T.-M. Yan, H.-Y. Cheng, C.-Y. Cheung, G.-L. Lin, Y.C. Lin and H.-L. Yu,
Phys. Rev. D 46, 1148 (1992).
 
\bibitem{georgi}
H. Georgi, Nucl. Phys. B 348, 293 (1991).

\bibitem{shifman1}
For a review see: {\it Vacuum structure and QCD sum rules}, edited by
 M.A. Shifman, North Holland, Amsterdam, 1992.

\bibitem{rad}
S.V. Mikhailov and A.V. Radyushkin, Phys. Rev. D 45, 1754 (1992).
 
\bibitem{neubert1} 
A.V. Radyushkin, Phys. Lett. B 271, 218 (1991);
M. Neubert, Phys. Rev. D 45, 2451 (1991).

\end{references}
\end{document}